\documentclass[aps,pra,reprint,groupedaddress,showpacs]{revtex4-1}
\usepackage{amssymb}
\usepackage{amsmath}
\usepackage{dcolumn}
\usepackage{bm}
\usepackage{graphicx}
\usepackage{mathrsfs}

\begin{document}

\title{Dissipative preparation of entangled states between two spatially separated nitrogen-vacancy centers}
\author{Peng-Bo Li}
\email{lipengbo@mail.xjtu.edu.cn}
\author{Shao-Yan Gao}
\email{gaosy@mail.xjtu.edu.cn}
\author{Hong-Rong Li}
\author{Sheng-Li Ma}
\author{Fu-Li Li}
\affiliation {MOE Key Laboratory for Nonequilibrium Synthesis and Modulation of Condensed Matter,\\
Department of Applied Physics, Xi'an Jiaotong University, Xi'an
710049, China}

\begin{abstract}
We present a novel scheme for the generation of entangled states of two spatially separated nitrogen-vacancy (NV) centers
with two whispering-gallery-mode (WGM)
microresonators, which are coupled either by an optical fiber-taper waveguide, or by the evanescent fields of the WGM. We show that, the steady state of the two NV centers can be steered into a
singlet-like state through a dissipative quantum dynamical process, where the cavity decay plays a positive role and can help drive the system to the target state. The protocol may open up promising perspectives for quantum communication and computation with this solid-state cavity quantum
electrodynamic system.

\end{abstract}
\maketitle

\section{Introduction}
Quantum entanglement is among the most fascinating aspects of
quantum mechanics, and entangled states of matter are now widely used
for fundamental tests of quantum theory and applications
in quantum information science \cite{quantum_information}. Plenty of different systems have been investigated to faithfully and controllably prepare entangled states of matter, among which
cavity QED \cite{Sci298,Kimble} offers one of the most promising and qualified candidates
for quantum state engineering and quantum information
processing. The general approach for quantum state engineering in cavity QED is based on either unitary dynamical evolution \cite{prl-75-3788,prl-89-187903,prl-85-1762,prl-91-097905,
prl-90-027903,prl-96-010503,OE-19-1207,OE-20-3176,pra-83-042302,pra-76-062311} or dissipative quantum dynamical process \cite{prl-90-047905,prl-106-090502,prl-86-4988,prl-102-073008,eprint,eprint1,pra-78-042307,pra-82-054103,pra-83-042329,pra-83-052312,pra-84-022316,pra-85-022324,pra-85-023812}. For the latter approach, the interaction of the
system with the environment is employed such that dissipation
drives the system into the desired state.  This process can be achieved by engineering the dissipative dynamics such that the desired state is the only steady state regardless of the initial state.

Recently, the solid-state counterpart of cavity QED system consisting of NV centers in diamond and WGM microresonator has attracted great interests \cite{NL-6-2075,OE-17-8081}, which circumvents
the complexity of trapping single atoms
and can potentially enable scalable device
fabrications. This composite system takes the advantage of both sides of NV centers and WGM microresonators, i.e., the exceptional spin properties of
NV centers \cite{Naure-Mat} and the ultrahigh quality factor and small mode volume of WGM microresonators \cite{prl-91-043902, nature-424-839,prl-105-153902,pra-57-R2293}. Therefore, the application of this solid state cavity QED system in quantum state engineering and quantum information
processing is of great interests \cite{nature-464-45,Proc-SPIE-6903-69030M,pra-83-054306,APL-96-241113,pra-83-054305,pra-84-011805,pra-84-032317,pra-84-043849}. In particular, it has interesting applications in  quantum networking  and quantum communication \cite{nature-453-1023}, since NV centers coupled to WGM microresonators are the natural candidates for quantum nodes, and these nodes can be connected by quantum channels such as optical fiber-taper waveguide.

In this work, we propose an efficient scheme for the preparation of entangled states of two spatially
separated NV centers with two microsphere resonators (MRs) coupled either by an
optical fiber-taper waveguide, or via the evanescent fields of the WGM. This proposal actively exploits the resonator decay to drive the system to a singlet-like entangled stationary state through an engineered reservoir for the NV centers.
We show that, the steady state of the two NV centers can be steered into a
singlet-like state through a dissipative quantum dynamical process. Up to our knowledge, this is the first proposal for preparing entanglement of distant NV centers employing  reservoir  engineering. Compared to previous works using unitary dynamical evolution \cite{pra-83-054306,APL-96-241113}, the present work has the following distinct features: (i) it performs well starting from
an arbitrary initial state, which renders the initialization
of the system in a pure state unnecessary; (ii) the produced entangled state is a pure stationary state; this auspicious
feature is very promising in view of the quest for viable, long-lived entanglement; (iii) since cavity decay has been used to actively drive the system dynamics, it thus converts a detrimental source of noise into a resource. This work may
represent promising steps toward the realization of entanglement with the solid-state cavity QED system.

\section{Dissipative entangled state preparation via reservoir engineering}
\subsection{Two WGM microresonators connected by an optical fiber-taper waveguide}
We first consider two negatively charged NV centers coupled to two remote MRs connected by a fiber-taper waveguide, as shown in Fig. 1. NV centers in diamond consist of a substitutional nitrogen atom and an adjacent vacancy having trapped an additional
electron, whose electronic ground state has a spin $S=1$ and is labeled as $\vert^3A_2\rangle=\vert E_0\rangle\otimes\vert m_s=0,\pm1\rangle$, where $\vert E_0\rangle$ is the orbital state with zero angular momentum projection along the NV axis. Quantum information is encoded in the spin states $\vert m_s=\pm1\rangle$ of the $^3A_2$ triplet such that $\vert0\rangle=\vert m_s=-1\rangle$, and $\vert1\rangle=\vert m_s=+1\rangle$. The $\Lambda$ three-level system could be realized in the NV center if the excited state $\vert e\rangle$ is chosen as $\vert e\rangle=\frac{1}{\sqrt{2}}(\vert E_-\rangle\vert m_s=+1\rangle+\vert E_+\rangle\vert m_s=-1\rangle)$ \cite{nature-466-730,pra-83-054306}, where $\vert E_{\pm}\rangle$ are orbital states with angular momentum projection $\pm1$
along the NV axis.

\begin{figure}[h]
\centerline{\includegraphics[bb=78 273 552 763,totalheight=3.5in,clip]{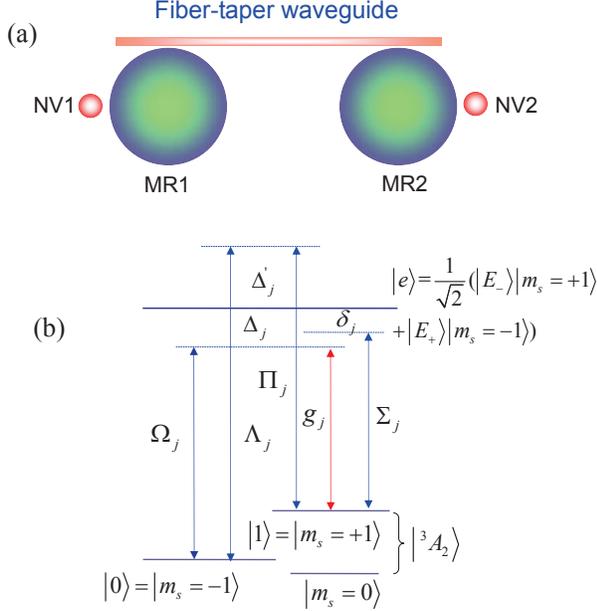}}
\caption{(Color online) (a) The schematic of two  NV centers coupled to two MRs, respectively, which are connected by a
fiber-taper waveguide. (b) Energy level structure with couplings to the cavity mode and driving laser fields. Quantum information is encoded in the spin states $\vert m_s=\pm1\rangle$ of the $^3A_2$ triplet, i.e., $\vert0\rangle=\vert m_s=-1\rangle$, and $\vert1\rangle=\vert m_s=+1\rangle$.
The excited state is $\vert e\rangle=\frac{1}{\sqrt{2}}(\vert E_-\rangle\vert m_s=+1\rangle+\vert E_+\rangle\vert m_s=-1\rangle)$.}
\end{figure}
The modes
of spherical resonators can be classified by mode numbers $n,
l$ and $m$, which determine the characteristic radial ($n$) and
angular ($l$ and $m$) field distribution of the modes. Usually the  so-called
fundamental WGM ($n = 1,l = m$) attracts great interests, whose field is
concentrated in the vicinity of the equatorial plane of the
sphere. In perfect spheres, resonance frequencies
do not depend on $m$, which means the MR supports both clockwise and counter-clockwise fundamental WGM. However,
it has been shown that this degeneracy can be lifted either by shape effect or by internal backscattering \cite{ol-20-1835}, and
these modes can be selectively excited by coupling to a tapered fiber \cite{prl-91-043902}. Therefore, in this work we only consider
a single fundamental WGM with frequency $\nu_0$ interacting with the NV centers.

The fundamental WGM dispersively
couples the transition $\vert 1\rangle\leftrightarrow\vert e\rangle$ for each N-V center with
coupling constant $g_j$.  The laser field of frequency $\omega_0$ couples the transition $\vert 0\rangle\leftrightarrow\vert e\rangle$ with Rabi frequency $\Omega_j$. These two transitions are assumed to be on the two-photon Raman resonance, thus establishing Raman transitions between the states $\vert 0\rangle$ and $\vert 1\rangle$ through the WGM and laser field. The laser fields of frequencies
$\omega_1$ and $\omega_2$ drive the transitions $\vert0\rangle\leftrightarrow\vert e\rangle$ and $\vert1\rangle\leftrightarrow\vert e\rangle$ with Rabi frequencies $\Lambda_j$ and $\Pi_j$, which are also on Raman resonance. Therefore, another Raman laser system is established between the states
$\vert 0\rangle$ and $\vert 1\rangle$ involving the two laser fields with frequencies $\omega_1$ and $\omega_2$.
The transition  $\vert1\rangle\leftrightarrow\vert e\rangle$ is also driven by a third classical laser field of frequency $\omega_3$ with Rabi frequency $\Sigma_j$. This laser field is used to induce an extra Stark shift for the state $\vert 1\rangle$, which can break the symmetry of the system Hamiltonian to ensure that
the system has a unique steady state \cite{eprint}. The corresponding detunings for the related transitions are $\Delta_j=\omega_{e0}-\omega_0=\omega_{e1}-\nu_0,\Delta'_j=\omega_{e0}-\omega_1=\omega_{e1}-\omega_{2},\delta_j=\omega_{e1}-\omega_{3}$, where $\omega_{e0},\omega_{e1}$ are the transition frequencies for the NV centers.
In the interaction
picture, the Hamiltonian describing the interaction between the NV centers and the laser fields and cavity modes
is (let $\hbar=1$)
\begin{eqnarray}
\label{H1}
\mathscr{H}_1
&=&\sum_{j=1,2}[\Omega_j\vert e\rangle_j\langle 0\vert e^{i\Delta_j t}+\Lambda_j\vert e\rangle_j\langle 0\vert e^{i\Delta'_jt}\nonumber\\
&&+\Pi_j\vert e\rangle_j\langle 1\vert e^{i\Delta'_jt}+\Sigma_j\vert e\rangle_j\langle 1\vert e^{i\delta_jt}+g_j\hat{a}_j\vert e\rangle_j\langle 1\vert e^{i\Delta_jt}]+\mbox{H.c.},\nonumber\\
\end{eqnarray}
where $\hat{a}_j$ is the annihilation operator for the j\emph{th} cavity mode.

We now discuss the coupling between the two MRs and the optical fiber-taper waveguide. We consider an optical fiber-taper waveguide
near the equatorial planes of both microspheres \cite{nature-424-839}, with
length $l$ and the decay rate of the resonator' fields into the continuum
of the fiber modes $\tilde{\nu}$. The number of longitudinal modes
of the fiber that significantly interact with the corresponding
resonator modes is on the order of $l\bar{\nu}/2\pi c$ \cite{prl-96-010503,OE-19-1207,pra-83-042302,pra-76-062311}. If we consider the short fiber limit $l\bar{\nu}/2\pi c\leq1$, then only one resonant mode $\hat{b}$ of the
fiber interacts with the cavity modes. Then the interaction Hamiltonian describing the coupling between the cavity modes and the fiber mode is \cite{prl-96-010503,OE-19-1207,pra-83-042302,pra-76-062311}
\begin{eqnarray}
\label{H2}
\mathscr{H}_{c,f}&=&\nu \hat{b}(\hat{a}_1^\dag+e^{i\varphi}\hat{a}_2^\dag)+\mbox{H.c.},
\end{eqnarray}
where $\nu$ is the cavity-fiber
coupling strength, and $\varphi$ is the phase due to propagation
of the field through the fiber. Define three normal bosonic modes $\hat{c},\hat{c}_1,\hat{c}_2$ by the canonical transformations
$\hat{c}=\frac{1}{\sqrt{2}}(\hat{a}_1-e^{-i\varphi}\hat{a}_2),\hat{c}_1=\frac{1}{2}(\hat{a}_1+e^{-i\varphi}\hat{a}_2+\sqrt{2}\hat{b}),\hat{c}_2=\frac{1}{2}(\hat{a}_1+e^{-i\varphi}\hat{a}_2-\sqrt{2}\hat{b})$. In terms of the bosonic modes $\hat{c},\hat{c}_1$ and $\hat{c}_2$, the interaction Hamiltonian $\mathscr{H}_{c,f}$ is diagonal. We rewrite this Hamiltonian as $\mathscr{H}_0=\sqrt{2}\nu \hat{c}_1^\dag \hat{c}_1-\sqrt{2}\nu \hat{c}_2^\dag \hat{c}_2$. So the whole Hamiltonian
in the interaction picture is $\mathscr{H}=\mathscr{H}_0+\mathscr{H}_1$.

In the following, the system-environment interaction is assumed Markovian, and  then is described
by a master equation in Lindblad form
\begin{eqnarray}
\label{M1}
\frac{d\rho}{dt}&=&i[\rho,\mathscr{H}]+\mathscr{D}_{R_1}\rho+\mathscr{D}_{R_2}\rho+\mathscr{D}_f\rho+\mathscr{D}_{\text{spon}}\rho,
\end{eqnarray}
with
\begin{eqnarray}
\mathscr{D}_{Rj}\rho&=&\kappa_j(2\hat{a}_j\rho\hat{a}^\dag_j-\hat{a}^\dag_j\hat{a}_j\rho-\rho\hat{a}^\dag_j\hat{a}_j),\nonumber\\
\mathscr{D}_f\rho&=&\kappa_f(2\hat{b}\rho\hat{b}^\dag-\hat{b}^\dag\hat{b}\rho-\rho\hat{b}^\dag\hat{b}),
\end{eqnarray}
where $\kappa_j$ is the leakage rate of photons from resonator $j$ ($\kappa_1=\kappa_2=\kappa$ is assumed in the following) and $\kappa_f$
is the decay rate of the fiber mode. The term $\mathscr{D}_{\text{spon}}\rho$ describes spontaneous emission of the NV centers from the excited state $\vert e\rangle$. Its concrete form is not relevant to this proposal, since we will adiabatically eliminate the excited state in the following.

We proceed to perform the unitary transformation $e^{-i\mathscr{H}_0t}$, which leads to
\begin{eqnarray}
\label{H3}
\mathscr{H}&=&\sum_{j=1,2}[\Omega_j\vert e\rangle_j\langle 0\vert e^{i\Delta_j t}+\Lambda_j\vert e\rangle_j\langle 0\vert e^{i\Delta'_jt}\nonumber\\
&&+\Pi_j\vert e\rangle_j\langle 1\vert e^{i\Delta'_jt}+\Sigma_j\vert e\rangle_j\langle 1\vert e^{i\delta_jt}]\nonumber\\
&&+\frac{1}{2}g_1\vert e\rangle_1\langle1\vert e^{i\Delta_1t}(\hat{c}_1e^{-i\sqrt{2}\nu t}+\hat{c}_2e^{i\sqrt{2}\nu t}+\sqrt{2}\hat{c})\nonumber\\
&&+\frac{1}{2}g_2\vert e\rangle_2\langle1\vert e^{i\Delta_2t}(\hat{c}_1e^{-i\sqrt{2}\nu t}+\hat{c}_2e^{i\sqrt{2}\nu t}-\sqrt{2}\hat{c})+\mbox{H.c.}\nonumber\\
\end{eqnarray}
If we assume $\nu\gg\{\Delta_j,\Delta'_j,\delta_j,g_j\}$, we can safely neglect the nonresonant modes $\hat{c}_1,\hat{c}_2$ under the rotating-wave approximation.
Thus we obtain the following Hamiltonian
\begin{eqnarray}
\label{H4}
\mathscr{H}&=&\sum_{j=1,2}\Omega_j\vert e\rangle_j\langle 0\vert e^{i\Delta_j t}+\Lambda_j\vert e\rangle_j\langle 0\vert e^{i\Delta'_jt}\nonumber\\
&&+\Pi_j\vert e\rangle_j\langle 1\vert e^{i\Delta'_jt}+\Sigma_j\vert e\rangle_j\langle 1\vert e^{i\delta_jt}]\nonumber\\
&&+\frac{1}{\sqrt{2}}g_1\vert e\rangle_1\langle1\vert e^{i\Delta_1t}\hat{c}-\frac{1}{\sqrt{2}}g_2\vert e\rangle_2\langle1\vert e^{i\Delta_2t}\hat{c}+\mbox{H.c.}\nonumber\\
\end{eqnarray}
In the large detuning limit, $\{|\Delta_j|,|\Delta'_j|,|\delta_j|,|\Delta_j-\Delta'_j|,|\Delta_j-\delta_j|,|\delta_j-\Delta'_j|\}\gg\{|\Omega_j|,|\Lambda_j|,|\Pi_j|,|g_j|\}$, we can adiabatically eliminate the excited state $\vert e\rangle$, and get an effective Hamiltonian describing two distinct Raman excitations
\begin{eqnarray}
\label{H5}
\mathscr{H}&=&-[\tilde{\Delta}_1\vert 1\rangle_1\langle 1\vert+\tilde{\Delta}_2\vert 1\rangle_2\langle 1\vert]
-[\Theta\vert 1\rangle_1\langle 0\vert\nonumber\\
 &&+\Theta\vert 1\rangle_2\langle 0\vert+g_{\text{eff}}\hat{c}^\dag\vert 1\rangle_1\langle 0\vert+g_{\text{eff}}\hat{c}^\dag\vert 1\rangle_2\langle 0\vert+\mbox{H.c.}],\nonumber\\
\end{eqnarray}
where $\tilde{\Delta}_j=\frac{|\Sigma_j|^2}{\delta_j}$, $\Theta=\frac{\Lambda_1\Pi_1^*}{\Delta'_1}=\frac{\Lambda_2\Pi_2^*}{\Delta'_2}$, and $g_{\text{eff}}=\frac{\Omega_1g_1^*}{\sqrt{2}\Delta_1}=-\frac{\Omega_2g_2^*}{\sqrt{2}\Delta_2}$.
In derivation of Hamiltonian (\ref{H5}), the resonance condition  $\frac{|\Omega_j|^2}{\Delta_j}+\frac{|\Lambda_j|^2}{\Delta'_j}\simeq\frac{|\Pi_j|^2}{\Delta'_j}+\frac{|g_j|^2}{2\Delta_j}\langle\hat{c}^\dag\hat{c}\rangle$ is used, and the constant terms has been omitted.

We now consider the dissipation term $\mathscr{D}_{R_1}\rho+\mathscr{D}_{R_2}\rho+\mathscr{D}_f\rho$, which can be simplified if we
give out its expression in terms of the bosonic modes $\hat{c},\hat{c}_1$ and $\hat{c}_2$.
From the canonical transformations
$\hat{b}=\frac{1}{\sqrt{2}}(\hat{c}_1-\hat{c}_2),\hat{a}_1=\frac{1}{2}(\hat{c}_1+\hat{c}_2+\sqrt{2}\hat{c}),\hat{a}_2=\frac{1}{2}e^{i\varphi}(\hat{c}_1+\hat{c}_2-\sqrt{2}\hat{c})$, we have
\begin{eqnarray}
   \mathscr{D}_{R_1}\rho&=& \kappa(2\hat{a}_1\rho\hat{a}^\dag_1-\hat{a}^\dag_1\hat{a}_1\rho-\rho\hat{a}^\dag_1\hat{a}_1) \nonumber\\
   &=&\frac{\kappa}{2}(\hat{c}_1+\hat{c}_2+\sqrt{2}\hat{c})\rho(\hat{c}_1^\dag+\hat{c}_2^\dag+\sqrt{2}\hat{c}^\dag)\nonumber\\
   &&-\frac{\kappa}{4}(\hat{c}_1^\dag+\hat{c}_2^\dag+\sqrt{2}\hat{c}^\dag)(\hat{c}_1+\hat{c}_2+\sqrt{2}\hat{c})\rho\nonumber\\
   &&-\frac{\kappa}{4}\rho(\hat{c}_1^\dag+\hat{c}_2^\dag+\sqrt{2}\hat{c}^\dag)(\hat{c}_1+\hat{c}_2+\sqrt{2}\hat{c}).
\end{eqnarray}
Since under the rotating-wave approximation the nonresonant modes $\hat{c}_1,\hat{c}_2$ are mostly in the vacuum state and have been neglected, we can discard the terms
$\hat{c}_i\rho\hat{c}_j^\dag,\hat{c}_i^\dag\hat{c}_j\rho,\rho\hat{c}_i^\dag\hat{c}_j,\hat{c}_i\rho\hat{c}^\dag,
\hat{c}\rho\hat{c}_i^\dag,\hat{c}_i^\dag\hat{c}\rho,\hat{c}^\dag\hat{c}_i\rho,\rho\hat{c}_i^\dag\hat{c},\rho\hat{c}^\dag\hat{c}_i$.
The same considerations can be applied to $\mathscr{D}_{R_2}\rho$ and $\mathscr{D}_f\rho$. In this case, $\mathscr{D}_{R1}\rho+\mathscr{D}_{R2}\rho+\mathscr{D}_f\rho$ can be approximated as \cite{pra-76-062311}
\begin{eqnarray}
\mathscr{D}_R\rho&=&\kappa(2\hat{c}\rho\hat{c}^\dag-\hat{c}^\dag\hat{c}\rho-\rho\hat{c}^\dag\hat{c}).
\end{eqnarray}

Now, the master equation (\ref{M1}) reduces to
\begin{eqnarray}
\label{M2}
\frac{d\rho}{dt}&=&i[\rho,\mathscr{H}]+\mathscr{D}_{R}\rho¡£
\end{eqnarray}
We can introduce the photon number representation for the density operator $\rho$ with respect to the normal mode $\hat{c}$, i.e., $\rho=\sum_{m,n=0}^\infty\rho_{mn}\vert m\rangle\langle n\vert$, where $\rho_{mn}$ are the field-matrix elements in the basis of the photon number states of the normal mode $\hat{c}$, and are still operators with respect to the NV centers. We now assume that the resonator modes are strongly damped, in which case the populations of the highly excited modes can be neglected. Therefore, we consider only the matrix elements $\rho_{mn}$ inside the subspace $\{\vert 0\rangle, \vert 1\rangle\}$ of the photon numbers. Under this approximation, the master equation (\ref{M2}) leads to the following set of coupled equations of motion for the field-matrix elements \cite{pr-372-369}
\begin{subequations}
\begin{eqnarray}
\dot{\rho}_{00}&=&\hat{F}\rho_{00}-ig_{\text{eff}}\rho_{01}\vert1\rangle_1\langle0\vert-ig_{\text{eff}}\rho_{01}\vert1\rangle_2\langle0\vert\nonumber\\
&&+ig_{\text{eff}}\vert0\rangle_1\langle1\vert\rho_{10}
+ig_{\text{eff}}\vert0\rangle_2\langle1\vert\rho_{10}+\kappa\rho_{11},
\end{eqnarray}
\begin{eqnarray}
\dot{\rho}_{11}&=&\hat{F}\rho_{11}-ig_{\text{eff}}\rho_{10}\vert0\rangle_1\langle1\vert-ig_{\text{eff}}\rho_{10}\vert0\rangle_2\langle1\vert\nonumber\\
&&+ig_{\text{eff}}\vert1\rangle_1\langle0\vert\rho_{01}
+ig_{\text{eff}}\vert1\rangle_2\langle0\vert\rho_{01}-\kappa\rho_{11},\nonumber\\
\end{eqnarray}
\begin{eqnarray}
\dot{\rho}_{01}&=&\hat{F}\rho_{01}-ig_{\text{eff}}\rho_{00}\vert0\rangle_1\langle1\vert-ig_{\text{eff}}\rho_{00}\vert0\rangle_2\langle1\vert\nonumber\\
&&+ig_{\text{eff}}\vert0\rangle_1\langle1\vert\rho_{11}
+ig_{\text{eff}}\vert0\rangle_2\langle1\vert\rho_{11}
-\frac{\kappa}{2}\rho_{01},\nonumber\\
\end{eqnarray}
\end{subequations}
where
\begin{eqnarray}
\hat{F}\rho_{ij}&=&-i[\mathscr{H}_d,\rho_{ij}]\nonumber\\
&=&i\Theta[\vert 1\rangle_1\langle 0\vert+\vert 0\rangle_1\langle 1\vert+\vert 1\rangle_2\langle 0\vert+\vert 0\rangle_2\langle 1\vert,\rho_{ij}]\nonumber\\
&&+i[\tilde{\Delta}_1\vert 1\rangle_1\langle 1\vert+\tilde{\Delta}_2\vert 1\rangle_2\langle 1\vert,\rho_{ij}]
\end{eqnarray}

The reduced density operator for the NV  centers is $\varrho_N=\mbox{Tr}_F(\rho)\simeq\rho_{00}+\rho_{11}$. For the case of strong resonator damping, we can adiabatically eliminate the elements $\rho_{01}$ and $\rho_{11}$ from the above equations \cite{pr-372-369}, prompting the evolution of the two NV centers with an effective master equation
\begin{eqnarray}
\label{M3}
\frac{d\varrho_N}{dt}&=&i[\varrho_N,\mathscr{H}_d]+\mathscr{L}_{e}\varrho_N\mathscr{L}_{e}^\dag-\frac{1}{2}(\mathscr{L}_{e}^\dag\mathscr{L}_{e}\varrho_N+\varrho_N\mathscr{L}_{e}^\dag\mathscr{L}_{e}),\nonumber\\
\end{eqnarray}
where
\begin{eqnarray}
\mathscr{H}_d&=&-[\tilde{\Delta}_1\vert 1\rangle_1\langle 1\vert+\tilde{\Delta}_2\vert 1\rangle_2\langle 1\vert
+\Theta\vert 1\rangle_1\langle 0\vert+\Theta\vert 1\rangle_2\langle 0\vert\nonumber\\
&&+\Theta\vert 0\rangle_1\langle 1\vert+\Theta\vert 0\rangle_2\langle 1\vert]\\
\mathscr{L}_{e}&=&\sqrt{\frac{4g^2_{\text{eff}}}{\kappa}}(\vert 1\rangle_1\langle0\vert+\vert 1\rangle_2\langle0\vert).
\end{eqnarray}
The first term in the right side of master equation (\ref{M3}) describes the coherent laser driving of the two NV centers, while the last two terms describe an effective engineered reservoir for the NV centers.

In order to gain more insight into the combined effect of the unitary and dissipative dynamics described by the master equation (\ref{M3}), we switch to the collective state picture for the two NV centers, with the three triplet states $\{\vert00\rangle=\vert0\rangle_1\vert0\rangle_2,\vert11\rangle=\vert1\rangle_1\vert1\rangle_2,\vert T\rangle=1/\sqrt{2}(\vert0\rangle_1\vert1\rangle_2+\vert1\rangle_1\vert0\rangle_2)\}$, and the singlet state $\vert S\rangle=1/\sqrt{2}(\vert0\rangle_1\vert1\rangle_2-\vert1\rangle_1\vert0\rangle_2)$. In this case, the  Hamiltonian $\mathscr{H}_d$ and the Lindblad operator $\mathscr{L}_{e}$ can be rewritten as
\begin{eqnarray}
\mathscr{H}_d&=&-[\sqrt{2}\Theta\vert11\rangle\langle T\vert+\sqrt{2}\Theta\vert T\rangle\langle 00\vert-\tilde{\Delta}\vert S\rangle\langle T\vert+\mbox{H.c.}]\\
\mathscr{L}_{e}&=&\sqrt{\frac{8g^2_{\text{eff}}}{\kappa}}(\vert 11\rangle\langle T\vert+\vert T\rangle\langle 00\vert),
\end{eqnarray}
where we have taken $\tilde{\Delta}_1=-\tilde{\Delta}_2=\tilde{\Delta}$.
The effective processes described by the Hamiltonian $\mathscr{H}_d$ and the Lindblad operator $\mathscr{L}_{e}$ is shown
in Fig. 2. The Hamiltonian $\mathscr{H}_d$ corresponding to  coherent laser driving  induces transitions between the three triplet states   $\{\vert00\rangle,\vert11\rangle,\vert T\rangle\}$, and between the triplet state$\vert T\rangle$ and the singlet state $\vert S\rangle$,  while the Lindblad operator $\mathscr{L}_{e}$ will drive the transition from $\vert 00\rangle$ to $\vert T\rangle$, and from $\vert T\rangle$ to $\vert 11\rangle$. Hence, the combined effect of the unitary and dissipative dynamics drives essentially much of the population to the singlet state $\vert S\rangle$, and a minor overlap with $\vert11\rangle$.
It can readily be seen that the state
\begin{eqnarray}
\vert \psi_S\rangle&=&\frac{1}{\sqrt{2\Theta^2+\tilde{\Delta}^2}}(\tilde{\Delta}\vert11\rangle+\sqrt{2}\Theta\vert S\rangle)
\end{eqnarray}
is the unique stationary state \cite{eprint} of the master equation (\ref{M3}) i.e.,
\begin{eqnarray}
\varrho_N(t\rightarrow\infty)&=&\vert \psi_S\rangle\langle \psi_S\vert
\end{eqnarray}
 Therefore, starting from an arbitrary initial state, we can prepare the steady state of the two NV centers in the singlet-like entangled state $\vert \psi_S\rangle$.
\begin{figure}[h]
\centerline{\includegraphics[bb=132 561 377 755,totalheight=2.1in,clip]{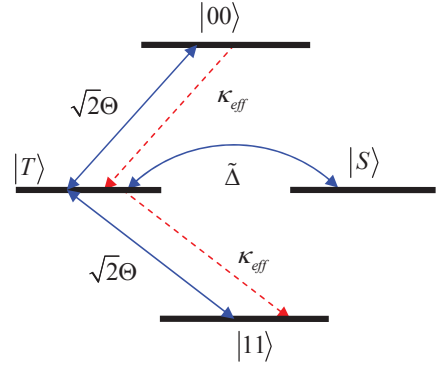}}
\caption{(Color online) Effective processes of the system. The driving $\sqrt{2}\Theta$ causes transitions between the three triplet states.
The NV centers decay through the cavities from $\vert00\rangle$ to $\vert T\rangle$, and from $\vert T\rangle$ to $\vert 11\rangle$ with
an effective decay rate $\kappa_{\text{eff}}=8g^2_{\text{eff}}/\kappa$. The singlet $\vert S\rangle$ is coherently
coupled to $\vert T\rangle$ by the level shift $\tilde{\Delta}$.}
\end{figure}
It should be noted that a similar master equation to (\ref{M3}) has been
presented in Refs. \cite{eprint,eprint1}, where they study the preparation of a maximally entangled state of two trapped atoms in a heavily damped cavity.
As pointed in Refs. \cite{eprint,eprint1}, to guarantee that the state $\vert \psi_S\rangle$ is the unique stationary state of the master equation (\ref{M3}),
breaking the symmetry in the system accomplished by a small energy level shift is necessary.

\subsection{Two coupled WGM microresonators through the evanescent fields of the WGM}
We point out that the essential idea of above discussions can be applied to the system in which the two WGM resonators couple with each other directly via
coherent photon hopping \cite{pra-79-042339,pra-84-013808}, i.e., coupled resonators. It can also be applied to the situation where two negatively charged N-V centers are  positioned
near the surface of a common WGM resonator \cite{pra-83-054306}. In the following we will discuss the former case in more detail.

\begin{figure}[h]
\centerline{\includegraphics[bb=169 620 490 804,totalheight=1.3in,clip]{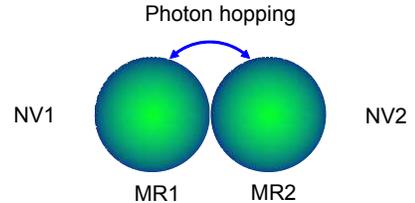}}
\caption{(Color online) The schematic of two  NV centers coupled to two MRs, respectively, which are coupled via the evanescent fields of the WGM}
\end{figure}
We now investigate the setup in which the two WGM resonators interact with each other directly via the evanescent fields of the WGM rather than an optical fiber-taper waveguide, as illustrated in Fig. 3. The couplings between the resonator modes and the laser fields are just the same as that shown in Fig. 1(b), except that the detuning for the WGM is now taken as $\bar{\Delta}_j$. In this case, the interaction between the NV centers and the laser fields and cavity modes is just described by the Hamiltonian (\ref{H1}). The Hamiltonian describing the direct coupling between these two resonators is
\begin{eqnarray}
\mathscr{H}_0&=&J(\hat{a}_1^\dag\hat{a}_2+\hat{a}_1\hat{a}_2^\dag),
\end{eqnarray}
where $J$ is the hopping rate of photons between the resonators. At present, the master equation describing the system-environment interaction is
\begin{eqnarray}
\label{M4}
\frac{d\rho}{dt}&=&i[\rho,\mathscr{H}]+\mathscr{D}_{R1}\rho+\mathscr{D}_{R2}\rho+\mathscr{D}_{\text{spon}}\rho,
\end{eqnarray}
where $\mathscr{H}=\mathscr{H}_0+\mathscr{H}_1$.

Introducing two normal modes $\hat{d}_1=1/\sqrt{2}(\hat{a}_1+\hat{a}_2)$, and $\hat{d}_2=1/\sqrt{2}(\hat{a}_1-\hat{a}_2)$, then the Hamiltonian $\mathscr{H}_0$ will have the diagonal form $\mathscr{H}_0=J(\hat{d}_1^\dag\hat{d}_1-\hat{d}_2^\dag\hat{d}_2)$.
Performing the unitary transformation $e^{-i\mathscr{H}_0t}$ will lead to
\begin{eqnarray}
\label{H6}
\mathscr{H}&=&\sum_{j=1,2}[\Omega_j\vert e\rangle_j\langle 0\vert e^{i\Delta_j t}+\Lambda_j\vert e\rangle_j\langle 0\vert e^{i\Delta'_jt}\nonumber\\
&&+\Pi_j\vert e\rangle_j\langle 1\vert e^{i\Delta'_jt}+\Sigma_j\vert e\rangle_j\langle 1\vert e^{i\delta_jt}]\nonumber\\
&&+\frac{1}{\sqrt{2}}g_1\vert e\rangle_1\langle1\vert e^{i\bar{\Delta}_1t}(\hat{d}_1e^{-iJ t}+\hat{d}_2e^{iJt})\nonumber\\
&&+\frac{1}{\sqrt{2}}g_2\vert e\rangle_2\langle1\vert e^{i\bar{\Delta}_2t}(\hat{d}_1e^{-iJ t}-\hat{d}_2e^{iJ t})+\mbox{H.c.}
\end{eqnarray}
Taking  $\bar{\Delta}_j-J=\Delta_j,|\bar{\Delta}_j+J|\gg \{|\Delta_j|,|\Delta'_j|,|\delta_j|,g_j\}$, this condition enables us to neglect the nonresonant mode $\hat{d}_2$  under the rotating-wave approximation. Consequently we obtain the following Hamiltonian
\begin{eqnarray}
\label{H7}
\mathscr{H}&=&\sum_{j=1,2}[\Omega_j\vert e\rangle_j\langle 0\vert e^{i\Delta_j t}+\Lambda_j\vert e\rangle_j\langle 0\vert e^{i\Delta'_jt}\nonumber\\
&&+\Pi_j\vert e\rangle_j\langle 1\vert e^{i\Delta'_jt}+\Sigma_j\vert e\rangle_j\langle 1\vert e^{i\delta_jt}]\nonumber\\
&&+\frac{1}{\sqrt{2}}g_1\vert e\rangle_1\langle1\vert e^{i\Delta_1t}\hat{d}_1+\frac{1}{\sqrt{2}}g_2\vert e\rangle_2\langle1\vert e^{i\Delta_2t}\hat{d}_1+\mbox{H.c.}\nonumber\\
\end{eqnarray}
If we further assume $\{|\Delta_j|,|\Delta'_j|,|\delta_j|,|\Delta_j-\Delta'_j|,|\Delta_j-\delta_j|,|\delta_j-\Delta'_j|\}\gg\{|\Omega_j|,|\Lambda_j|,|\Pi_j|,|g_j|\}$, then we can adiabatically eliminate the excited state $\vert e\rangle$, and get an effective Hamiltonian in a rotating frame as that of Eq. (\ref{H5})
\begin{eqnarray}
\label{H8}
\mathscr{H}&=&-[\tilde{\Delta}_1\vert 1\rangle_1\langle 1\vert+\tilde{\Delta}_2\vert 1\rangle_2\langle 1\vert]
-[\Theta\vert 1\rangle_1\langle 0\vert\nonumber\\
 &&+\Theta\vert 1\rangle_2\langle 0\vert+g_{\text{eff}}\hat{d}_1^\dag\vert 1\rangle_1\langle 0\vert+g_{\text{eff}}\hat{d}_1^\dag\vert 1\rangle_2\langle 0\vert+\mbox{H.c.}]\nonumber\\
\end{eqnarray}
Following the same reasoning as that for getting the master equation (\ref{M3}), we can get an effective master equation which has the same
form as Eq. (\ref{M3}).
In this case, we can also prepare the stationary state of the two NV centers in an entangled state.
\section{numerical simulations and practical considerations}

It is necessary to verify the model through numerical
simulations. To provide an example, we consider the tapered-fiber case. The same results can be obtained in
the case where the microresonators are directly coupled. Taking account of the dephasing effect, we simulate
the dynamics of the system by the following master equation
\begin{eqnarray}
\label{M5}
\frac{d\rho}{dt}&=&i[\rho,\mathscr{H}]+\mathscr{D}_{R_1}\rho+\mathscr{D}_{R_2}\rho+\mathscr{D}_f\rho+\mathscr{D}_{\text{deph}}\rho,
\end{eqnarray}
where
\begin{eqnarray}
\mathscr{D}_{\text{deph}}\rho&=&\gamma_\varphi\sum_{j=1}^2(2\sigma_z^j\rho\sigma_z^j-\sigma_z^j\sigma_z^j\rho-\rho\sigma_z^j\sigma_z^j)
\end{eqnarray}
with $\gamma_\varphi$ the pure dephasing rate of the NV centers, and $\sigma_z^j=\vert1\rangle_j\langle 1\vert-\vert0\rangle_j\langle 0\vert$.
In the following through solving the full master equation (\ref{M5}) numerically, we calculate the time evolution of the fidelity, and of the population of the singlet state and of the three triplet states for different initial state. The fidelity with respect to the singlet-like state $\vert \psi_S\rangle$ is defined as $F=\langle\psi_S\vert\varrho_N\vert\psi_S\rangle$.
In order to perform the simulation much easier, we switch to the field representation with respect to the normal modes $\hat{c},\hat{c}_1,\hat{c}_2$.
Then the Hamiltonian in equation (\ref{M5}) is taken as (\ref{H3}).
To solve the master equation numerically,
we employ the Monte Carlo wave function formalism from the quantum trajectory
method \cite{CPC}. All the simulations are performed under 50 trajectories.
The relevant parameters are chosen such that they are within the parameter
range for which the scheme is valid.

\begin{figure}[h]
\centerline{\includegraphics[bb=29 21 633 635,totalheight=2.5in,clip]{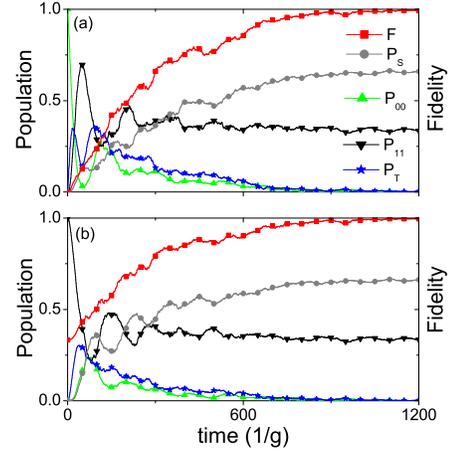}}
\caption{(Color online) Time evolution of the fidelity $F$, of the respective population of the singlet state and the three triplet states $P_S,P_{00},P_{11},P_{T}$ from two different initial states, (a) $\vert00\rangle$; (b) $\vert11\rangle$. The parameters are chosen as $g_1\sim g_2\simeq g,\nu\simeq 100g,\Delta_j\sim-\Delta_j'\simeq10g,\delta_1\sim-\delta_2\simeq20g,\Omega_1\sim-\Omega_2\simeq g,\Lambda_j\simeq g,\Pi_j\simeq 0.1g,\Sigma_j\simeq \sqrt{5}g/5,\kappa=0.5g,\gamma_\varphi=0$.}
\end{figure}
Figure 4 shows the numerical results for the fidelity $F$, the population of the singlet state $P_S$, and the respective population of the three triplet states $P_{00},P_{11},P_{T}$ for an appropriate set of parameters, starting from two different initial states $\vert00\rangle$ and $\vert11\rangle$  without dephasing.
One can readily see that the model performs very well with the chosen parameters. Starting from a given initial state, the populations $P_{00},P_{T}$ oscillate rapidly with an envelop decaying at a rate proportional to $g_{eff}^2/\kappa$, while the fidelity $F$ and the populations $P_{S}$ and $P_{11}$ converge to the maximum values at the same rate. In fact, the system can evolve to the stationary state $\vert\psi_S\rangle$ with a fidelity higher than $99\%$ regardless of the initial state. Figure 5 displays time evolution of the fidelity under several dephasing rates starting from
two different initial states $\vert00\rangle$ and $\vert11\rangle$. From the simulations we can see that, when $\gamma_\varphi<\{\Theta,g_{eff}^2/\kappa\}$, the effect of dephasing on the fidelity of the scheme can be neglected. However, as the dephasing rate $\gamma_\varphi$
becomes comparable to the effective coherent coupling strength $\Theta$, or larger than it, the fidelity of the scheme is spoiled by dephasing.
Therefore, to implement this proposal with high fidelity requires that $\gamma_\varphi<\{\Theta,g_{eff}^2/\kappa\}$.

\begin{figure}[h]
\centerline{\includegraphics[bb=29 21 633 635,totalheight=2.5in,clip]{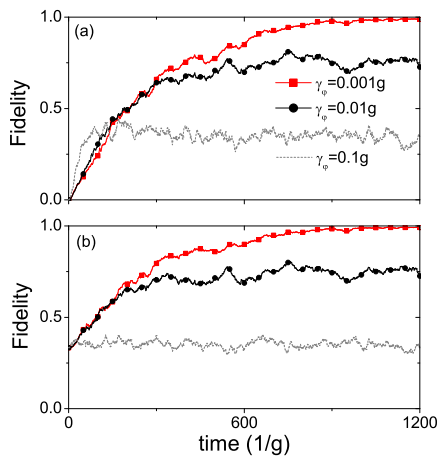}}
\caption{(Color online) Time evolution of the fidelity $F$ under several dephasing rates starting from two different initial states, (a) $\vert00\rangle$; (b) $\vert11\rangle$. Other parameters are the same as those in Fig. 4}
\end{figure}

For experimental implementation of this protocol, the present day achievements in the experiment with NV centers and WGM resonators can be used.
Coherent coupling between individual NV center in diamond and the WGM in microsphere or microdisk resonator has been reached \cite{NL-6-2075,OE-17-8081}. For example, in the experiment demonstrating the normal mode splitting with NV centers in diamond nanocrystals
and silica microspheres \cite{NL-6-2075}, the coupling strength between NV centers and the WGM is about $g/2\pi\sim55$ MHz, and the cavity decay rate is $\kappa/2\pi\sim50$ MHz. For the NV centers, the electron spin relaxation
time $T_1$ of diamond NV centers ranges from several milliseconds at room
temperature to seconds at cryogenic temperature. For our scheme, it should be implemented at cryogenic temperatures ranging from 6 to 12 K \cite{NL-6-2075}. The dephasing time $T_2$ induced by the fluctuations in the nuclear spin bath has the value of several microseconds in general, which can be increased to 2 milliseconds in ultrapure diamond \cite{Naure-Mat}. Therefore, the time for reaching the stationary state $\vert\psi_S\rangle$, which is about $1000/g\sim3 \mu$s, is shorter than the typical decoherence time for this system. The two resonators can be coupled either by an optical fiber-taper waveguide for the case of remote resonators, or via the evanescent fields of the WGM for the case of nearby resonators. As for the coherent coupling between the fiber and resonators,
a perfect fiber-cavity coupling
(with efficiency larger than $99.9\%$) can be realized by
fiber-taper coupling to high-Q silica microspheres \cite{prl-91-043902} or
microtoroidal cavities \cite{prl-102-083601}. Thus with present day techniques in solid-state cavity QED, this proposal could be realized.

\section{Conclusion}
We have presented a novel scheme for the preparation of  entangled states of two spatially
separated NV centers with two WGM resonators coupled either by an
optical fiber-taper waveguide, or via the evanescent fields of the WGM. The proposal actively exploits the resonator decay to drive the system to a singlet-like entangled stationary state through an engineered reservoir for the NV centers. This work may
represent promising steps toward the realization of entanglement with the solid-state cavity QED system.
\section*{ACKNOWLEDGMENTS}
P.B.L. acknowledges the enlightening discussions with  Y. F. Xiao and Z. Q. Yin.
This work is supported by the NNSF of China under
Grant Nos.11104215 and 11174233, the Special Prophase
Project in the National Basic Research Program of China under
Grant No. 2011CB311807, and the Research Fund for the
Doctoral Program of Higher Education of China under Grant
No. 20110201120035. S.-Y. Gao acknowledges financial
support from the Natural Science Basic Research Plan in the
Shaanxi Province of China (No. 2010JQ1004).

\providecommand{\noopsort}[1]{}\providecommand{\singleletter}[1]{#1}%

\end{document}